# Bidirectional and Turbulence-Resilient Fi-Wi-Fi Bridge

Florian Honz, and Bernhard Schrenk

*AIT Austrian Institute of Technology, Center for Digital Safety & Security / Security & Communication Technologies, 1210 Vienna, Austria.*
*Author e-mail address: florian.honz@ait.ac.at*

**Abstract:** We present a full-duplex 10Gb/s FSO bridge between two single-mode ports, utilizing centralized beamforming and simultaneous channel sounding. We further mitigate turbulence-induced fading through diversity reception enabled by wavelength-set coding.

## 1. Introduction

Optical wireless communication (OWC) offers a high bandwidth and license-free operation, even in environments that are contaminated by electro-magnetic interference. Apart from local access applications such as in-door LiFi or underwater communications, OWC has been adopted in free-space optical (FSO) links that serve as multi-Tb/s fiber-in-the-air backhaul between fixed stations [1]. FSO links further support access-grade 1-10 Gb/s pipes between mobile terminals [2] and enable $100^+$ Gb/s inter-space comms [3], including the first FSO-enabled streaming of an ultra-high definition cat video at up to 267 Mb/s from a deep-space probe that is 19 million miles away from Earth [4]. However, the practical implementation of terrestrial FSO systems is burdened by their susceptibility to atmospheric effects such as poor weather or turbulent air, and by the high degree of system complexity required to accommodate for sophisticated pointing, acquisition and beam tracking. As a response to these challenges, recent works have investigated integrated beamformers [5,6], including diversity concepts to battle signal fading due to turbulence [7].

This work advances our earlier Fi-Wi-Fi bridge [8] and demonstrates a bidirectional FSO link that connects two single-mode ports through focal plane array (FPA) beamformers located at the fiber tips, including centralized control, channel sounding and a reflective FSO configuration that supports full-duplex 10 Gb/s transmission through wavelength-reuse. We further prove turbulence mitigation through a graceful upgrade towards diversity reception.

## 2. Bidirectional Air Interface with Remote Beamforming Control

Our air interface builds on a FPA beamformer to simplify control and calibration when setting the direction of the emitted pencil beam as a function of the offset to the focal center of its collimation lens [5]. The FPA builds on 91 single-mode cores (MFD of 8.4 µm, crosstalk -47 dB) that are hexagonally arranged over the cross-section of its focal plane, with typically $N$ = 11 elements per axis. The antenna elements are pitched by $\pi$ = 35 µm and are fed by a photonic lantern to yield individual fiber ports towards its beamforming network. The fill factor for all cores over the focal plane is 5.9% or -12.2 dB. The FPA is completed by a 2" lens with a focus of $f$ = 100 mm, yielding a FSO beam diameter of ~28 mm. The field-of-view of $FoV = N \cdot \pi / f$ = 0.22° is sufficiently wide for point-to-point links. The angled (8°) polish of the focal plane enables bidirectional transmission. By using an optical switch (SWI) and a (cyclic) arrayed waveguide grating (AWG) as beamforming network at the FPAs at the head-end (HE) and tail-end (TE) FSO terminals (Fig. 1a), the HE can remotely set the antenna element at the TE through simple wavelength choice, thanks to the colorless transmission of its switched beamforming network.

The bidirectional FSO link (Fig. 1b) has been evaluated for three scenarios (Table 1) that investigate trade-offs between system complexity and performance. Scenario ❶ uses a C-band setup for half-duplex 10 Gb/s on-off keyed down- (DS) and upstream (US) transmission, together with periodic channel sounding. Scenario ❷ dedicates an entire spectral band to the DS (L-band), the US (C-band) and channel sounding (S-band) to facilitate full-duplex data transmission with simultaneous evaluation of the FSO link quality. Scenario ❸ simplifies this layout by exploiting wavelength-reuse for the purpose of full-duplex C-band DS/US transmission. For this, it uses a reduced DS extinction ratio (ER) of 3 dB to enable remodulation of the DS [9]. Channel sounding is continuously performed at the L-band.

The notion of simplified TE terminals is supported through remote control and remote modulation: For these, the TE-side beamformer is entirely orchestrated by the HE, while a polarization-insensitive EAM transmitter at the TE is remotely seeded by the HE. The signals are launched with a power of 11.9 and 9.4 dBm from the HE and TE, respectively. Reception-wise, the TE employs an APD+TIA receiver after blue/red (B/R) band splitting in case a dedicated DS waveband is employed (❷), while the preamplified signal before the reflective EAM is split by a 50/50 coupler towards a simple PIN+TIA receiver in case of wavelength re-use (❸) or half-duplex operation (❶).

Channel sounding provides information for centralized beamforming control. It is based on white-light injection at the TE, whose coupling-dependent signature is acquired by a spectral monitor (OSA) at the HE to determine the optimal pair of antenna elements at both FPAs. The DS and US seed wavelengths at the HE are then tuned accordingly.

Figure 2a presents an example for the coupling between all HE- and TE-side FPA antenna elements over a 6-m in-door FSO link. The elements at the TE are referred to as wavelength channel λ, corresponding to the inwards-out spiral-shaped allocation of 50-GHz spaced AWG channels to the cores of its photonic lantern. As exemplified in Fig. 2a, the launch of the HE signal at core 55 results in a maximal coupling of -13.2 dBm at core 16 of the TE, at λ = 1542.89 nm. The poor coupling marked by Ξ indicates the faulty adjacent port 24 at the TE-side beamformer.

Figure 2b presents the optical spectrum when no signal is present and white light is injected at the TE on all three (S, C, L) wavebands to sound the FSO channel. The information obtained from Fig. 2b resembles that from Fig. 2a. Moreover, the coupling efficiency is spectrally repeated in adjacent wavebands by virtue of the free spectral range (FSR) property of the AWG. With its FSR of 5.96 THz, the FSO link quality is imprinted in the S-, C- and L-bands. Since we only need one waveband for channel sounding, the other two are available for data transmission.

Figure 2c reports the HE-side acquisition of a C-band US signal at Λ = 1541.97 nm, which has been remotely seeded by the HE through re-use of the DS signal as the optical US carrier. To enable this seeding scheme, the FPA elements are required to feature a high optical return loss. As Fig. 2c shows, the reflected DS signature that is received at the US receiver at the HE, which has been plotted for a deliberately blocked FSO link, is sufficiently weak to maintain a high optical signal-to-reflection ratio (OSRR) of 23.4 dB for US reception under bidirectional operation.

## 3. Transmission Performance over Fi-Wi-Fi Link and Mitigation of Signal Fading due to Turbulent Air

Data transmission has been evaluated in terms of real-time BER (Fig. 3) as function of the received optical power (ROP). Figure 3a presents the 10 Gb/s DS BER. The C-band layout in scenarios ❶ and ❸ with EDFA+PIN receiver yields a reception sensitivity of -25.4 dBm (●) and -18.8 dBm (▲) at a BER of $10^{-10}$, respectively. The average penalty of 5.2 dB ($\delta$) agrees with the expected value of 4.8 dB due to a reduced eye opening for a DS ER of 3 dB [9] when implementing full-duplex transmission by means of wavelength-reuse (❸). For the L-band DS with APD receiver (❷), which benefits from the optical noise filtering of the AWG at the TE, the sensitivity is -28 dBm at a BER of $10^{-10}$ (◆). The 10 Gb/s continuous-wave seeded US (Fig. 3b) features a sensitivity of -23.2 dBm (■) at a BER of $10^{-10}$ (❶,❷) and a penalty 5.3 dB ($\xi$) is incurred at the FEC level of $10^{-3}$ when the DS is used as optical US carrier (▲) in scenario ❸. This penalty can be reduced when adopting DS suppression techniques for symmetric data rates [9].

The FSO setup has then been installed at a roof-top location (Fig. 1c) at which the FPAs bridge a distance of L = 63 m. The corresponding DS BER performance (❶) for continuous real-time testing over 90 minutes on a sunny (49 klux) autumn day (13.9°C) is reported in Fig. 3c. We found the vast majority of 97.3% of BER acquisitions to fall below $10^{-9}$. However, we also notice several excursions in the BER that go with a fast fading in ROP. This is attributed to windy conditions during the first part of the measurement. Figure 3d shows the correlation between the fading of the monitored signal envelope and the recorded ROP. The fast fluctuations cannot be compensated by means of beamforming control due to its slow response. This inevitably results in deep fading for the received signal.

A mitigation method for excessive signal fading has been investigated by devoting a specific set of wavelengths to the DS emission, which together with the λ-routing at the TE-side FPA results in diversity reception. This is sketched in Fig. 4a, where a lateral offset of the FSO beam is caused as the light beam passes through turbulent air. This offset causes the beam to couple to neighboring cores, which relate to specific wavelength channels λ of the AWG. If the wavelength set for DS emission is configured to match these surrounding FPA elements (i.e. AWG channels) of the element at the TE that is currently considered as that yielding the optimal coupling under clear-air conditions, the signal fading can be reduced: Diversity reception then fills the drop in signal power through incoherent beam combination across the wavelength set that had been jointly modulated with the same DS signal.

The effectiveness of this scheme has been evaluated for an in-door FSO channel that has been perturbed by a heat gun set to 600°C and aiming at the launched beam at the HE (Fig. 1b). Figure 4b presents the effect on the sounded channel, showing that the coupled power becomes unstable at previously optimal wavelength channels while the power spreads among adjacent FPA antenna elements. This is evidenced by the acquired 10 Gb/s DS waveform (Fig. 4c), which under clear air features a constant envelope and an open eye diagram. As the turbulence is introduced to the FSO channel, the DS shows strong signal fading and a closed eye. The spectral DS launch is then expanded to a set of 7 wavelengths to enable diversity reception. This reduces the signal fading and re-opens the eye. The corresponding BER performance (Fig. 4d) confirms the mitigation of the high error floor at >$10^{-3}$ (▲) for single-wavelength launch under turbulent conditions: For the wavelength set (●), 10 Gb/s DS transmission can be again accomplished. A penalty of 3.4 dB (τ) remains compared to clear-air conditions (■), yet at a power margin of 7.2 dB.

## 4. Conclusion

We have demonstrated full-duplex 10 Gb/s transmission over a bidirectionally operated point-to-point FSO bridge that interconnects two single-mode fiber ports through centralized beamforming and simultaneous channel sounding. Successful full-duplex operation has been demonstrated on a single optical carrier. By encoding data on a wavelength set, turbulences can be additionally mitigated at an acceptable residual performance penalty. An improvement of the fill factor for the utilized FPA is left for a further reduction of the fiber-to-fiber coupling loss over the Fi-Wi-Fi bridge.

***Acknowledgement***: *This work has received funding from the Smart Networks and Services Joint Undertaking (SNS JU) under the European Union's Horizon-Europe research and innovation programme under Grant Agreement No. 101139182.*

| FSO configuration | DS | US | Sounding |
|---|---|---|---|
| ❶ Half-duplex | C-band, ER > 10 dB | C-band | C-band, periodic |
| ❷ Full-duplex with paired spectrum | L-band, ER > 10 dB | C-band | S-band, continuous |
| ❸ Full-duplex through λ re-use | C-band, ER = 3 dB | C-band | L-band, continuous |

Table 1. FSO scenarios and use of optical spectrum.

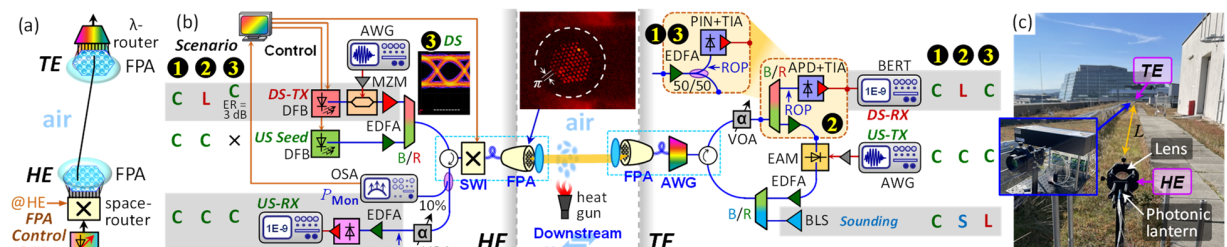

Fig. 1. (a) Fi-Wi-Fi bridge. (b) Experimental and spectral configuration of FSO link for investigated scenarios. The inset shows a focal-plane photograph of the FPA antenna elements at the fiber tip after illumination of the photonic lantern with red light. (c) Out-door FSO link setup.

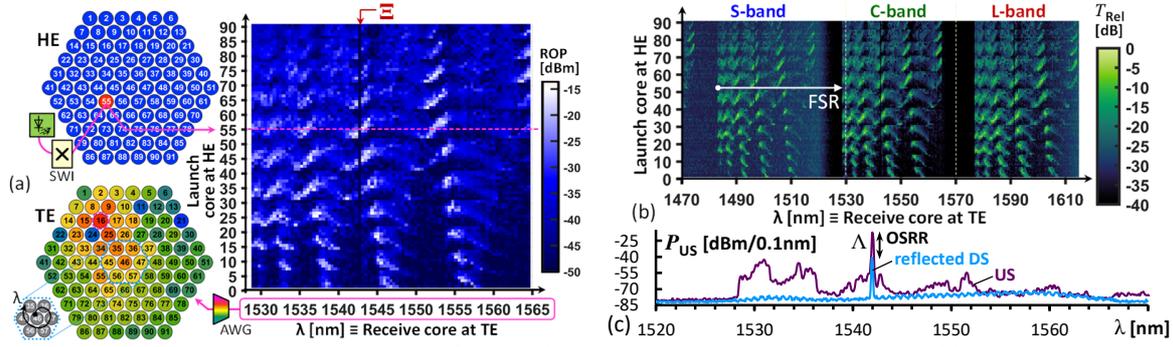

Fig. 2. (a) Coupling between HE- and TE-side FPA beamformers. (b) Channel sounding over multiple wavebands. (c) Received spectrum at HE.

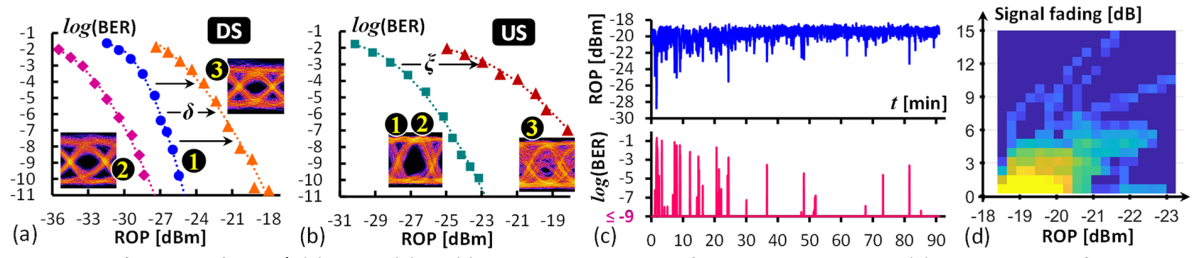

Fig. 3. BER performance of 10 Gb/s (a) DS and (b) US. (c) Long-term BER and ROP for out-door FSO link and (d) correlated signal fading.

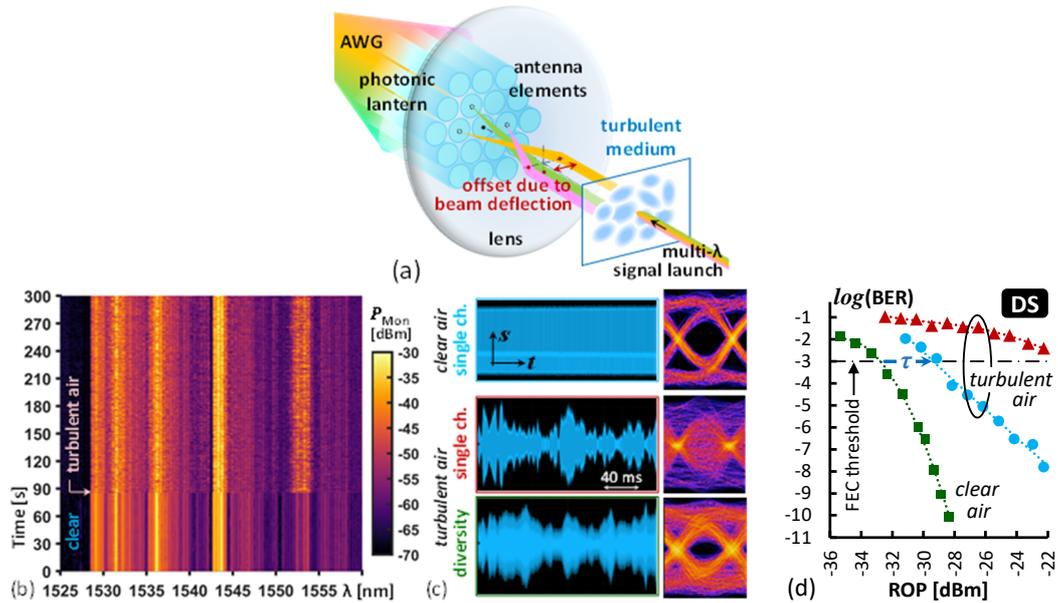

Fig. 4. (a) Diversity reception through multi-$\lambda$ launch. (b) Effect of turbulence on FSO coupling, (c) the signal envelope / eye, and (d) on the BER.